\documentclass [12pt]{iopart}
\usepackage{epsf}
\begin{document}
\title{Notes on static cylindrical shells}
\author{J Bi\v{c}\'{a}k and M \v{Z}ofka}
\address{Institute of Theoretical Physics, Faculty of Mathematics and Physics, Charles University Prague, V Hole\v{s}ovi\v{c}k\'{a}ch 2, 180 00 Praha 8, Czech Republic \\
\vspace{0.2cm} E-mail: \texttt{bicak@mbox.troja.mff.cuni.cz,
zofka@mbox.troja.mff.cuni.cz}}
\date{}
\begin{abstract}
Static cylindrical shells made of various types of matter are
studied as sources of the vacuum Levi-Civita metrics. Their
internal physical properties are related to the two essential
parameters of the metrics outside. The total mass per unit length
of the cylinders is always less than $\frac{1}{4}$. The results
are illustrated by a number of figures.
\end{abstract}
% \submitto{\CQG}
%\maketitle{}
%
\section{Introduction}
Cylindrically symmetric spacetimes and their sources have
belonged to useful models in classical relativity for many years.
Although recently attention is paid primarily to dynamical
situations in which gravitational waves are present (see, e.g.,
\cite{Bondi}, \cite{Goncalves} and references
therein) or to the cylindrical fields motivated by string theories (e.g. \cite{Horsky}), unsolved questions persist even in the standard static cases.

In contrast to the Schwarzschild metric described completely by
one parameter -- the Schwarzschild mass, the static vacuum
Levi-Civita (LC) solution contains two essential constant
parameters -- $m$, related to the local curvature, and $C$,
determining the global conicity of the spacetime. To connect these
two parameters with physical properties of cylindrical sources
turns out to be considerably more difficult than in the spherical
case. Here we do not wish to survey numerous contributions to the
subject -- we refer the reader to the recent review \cite{Bonnor};
some more references, in particular on static cylindrical shells,
are given in the following sections.

The purpose of these notes is to construct various types of
physically plausible cylindrical shell sources and to connect
their internal properties, especially their mass per unit length,
with the two parameters of the external LC metrics. We
consider not only the shells made of counter-rotating particles
with non-zero rest mass but also of photons. In addition, inspired
by recent work on two counter-moving light beams \cite{Gonna and
Kramer}, we study shells of counter-spiralling particles with both
non-zero and zero rest mass. Next, we discuss perfect fluids as well as shells made of anisotropic matter satisfying, respectively,
the weak, strong, and dominant energy conditions\footnote[1]{As
demonstrated recently \cite{Goncalves}, the particular type of the energy condition
plays an important role in the outcome of the gravitational
collapse of cylindrical shells. (See \cite{Goncalves} for a number of
references on dynamical cylindrical spacetimes.)}. We
determine permissible ranges of radii of the 2-dimensional
cylinders made of various types of matter. In all the cases
considered, we conclude that the mass per unit length based on the definition by Marder \cite{Marder} is restricted by $M_1 \leq \frac{1}{4}$. We
find that even for small values of the external LC parameter $m$,
it is important to realize that the conicity $C$ enters into
the relation between the internal parameter $M_1$ and the external
parameters (solving so a puzzle in literature). Although in most
of the text we assume flat spacetime inside the shells, we make
some remarks on the situations in which a 'cosmic string' is
located along the axis of the cylinder and on the limits that can
produce a cosmic string from a 'shrinking' cylinder.

At the end we add a few notes on the geodesics outside the
cylinder and, in Appendix, some basic properties of the Newtonian
cylindrical shells and the relativistic spherical shells are
summarized for comparison. In contrast to the literature on static
cylinders we are aware of, we also illustrate the results by a
number of figures.\\
\section{Cylindrical shells and their mass per unit length}
If the singularity is located on the axis of symmetry then the LC
metric in the Weyl coordinates $\tau \! \in \! I\hspace{-0.13cm}R,
r \! \in \! I\hspace{-0.13cm}R^+, \zeta \! \in \!
I\hspace{-0.13cm}R, \phi \! \in \! [\: 0,2\pi)$ can be written in
terms of two constant dimensionless parameters $m \! \in \!
I\hspace{-0.13cm}R$ and $C \! \in \! I\hspace{-0.13cm}R^+$ and
constant $r_0 \! \in \! I\hspace{-0.13cm}R^+$ with the dimension
of length as follows
\begin{equation}
ds^2 = -\left( \frac {r}{r_0} \right)^{2m} d\tau^2 +
\left( \frac {r}{r_0} \right)^{2m(m-1)} (d\zeta^2 + dr^2) + r^2
\left( \frac {r}{r_0} \right)^{-2m} \frac {d\varphi^2} {C^2}.
\end{equation}
Introducing dimensionless coordinates $\rho \equiv \frac {r}
{r_0}, \; t \equiv \frac {\tau} {r_0}, \; z \equiv \frac {\zeta}
{r_0},$ we obtain the metric in the standard dimensionless form
(see e.g. \cite{Bonnor}):

\begin{equation} \label{Coordinate System}
d \tilde {s} ^2 = ds^2 / r_0^2 = -\rho^{2m} dt^2 + \rho^{-2m}
\left[ \rho^{2m^2} (dz^2 + d\rho^2) + \frac {1} {C^2} \rho^2
d\varphi^2 \right].
\end{equation}

In general, we assume two LC solutions to be given in two
different sets of these coordinates, one solution inside the shell
($\rho \leq \rho_-$) and the other outside ($\rho \geq \rho_+$).
Using Israel's formalism \cite{Israel} we readily find expressions
for the energy-momentum tensor $S_{ij}$ ($i,j =
T,Z,\mathit{\Phi}$; the coordinates on the shell are chosen in
such a way that the induced 3-metric is flat) induced on the
shell:

\begin{eqnarray} \label{Induced_tensor}
8\pi S_{TT} = \rho_-^{m_--m_-^2-1} (1-m_-)^2 - \rho_+^{m_+-m_+^2-1} (1-m_+)^2, \nonumber\\
8\pi S_{ZZ} = \rho_+^{m_+-m_+^2-1} - \rho_-^{m_--m_-^2-1}, \\
8\pi S_{\mathit{\Phi}\mathit{\Phi}} = \rho_+^{m_+-m_+^2-1} m_+^2 -
\rho_-^{m_--m_-^2-1} m_-^2, \nonumber
\end{eqnarray}
with non-diagonal components vanishing. We require the spacetime
to be regular along the axis, thus we put\footnote[2]{If we wish
to admit a non-zero missing angle (a "cosmic string") inside the
shell we need to replace $C_+ \rightarrow C_+/C_-$ in the
subsequent formulae. This does not have any major consequences.}
$m_- \!\! = \! 0, \: C_- \! = \! 1$. The proper length of a
hoop with $t, \rho, z$ constant must be the same from both sides
of the cylinder:

\begin{equation}
2\pi \rho_- = 2\pi \rho_+^{1-m_+} /C_+.
\label{Hoops}
\end{equation}
We define the mass per unit (coordinate) length of the cylinder,
following \cite{Marder}, by

\begin{equation}
\label{M1 - definition} M_1 \equiv \mbox{(Circumference)} \cdot \;
S_{TT} = 2\pi \rho_- S_{TT},
\end{equation}
where $S_{TT}$ is given in equation (\ref{Induced_tensor}). Equations
(\ref{Induced_tensor}) and (\ref{M1 - definition}) yield

\begin{equation}
\label{M1} M_1 = \frac {1} {4} \left( 1- \frac {1} {C_+} \frac
{(1-m_+)^2} {\rho_+ ^{m_+^2}}\right).
\end{equation}
It is immediately seen that $M_1 \leq \frac {1} {4}$. In case of 3-dimensional solid cylinders, an analogous restriction has been recently shown to hold for energy density per unit proper length of the cylinder (see \cite {Anderson}, below equation (3.11)). Our result holds for {\it any} matter on the shell. It is quite in accord
with the notion that a spacetime without singularities is free of
horizons if the amount of matter within a given region is bounded
by a certain finite value. This limiting value can be exceeded
only at the expense of a singularity along the axis. Indeed, the
case of $M_1 \! > \! \frac{1}{4}$ requires $m_- \! \not = \! 0$,
causing a singularity along the symmetry axis. Admitting $m_- \not
= 0$, we find

\[\label{M1C_-} M_1 = \frac {1} {4} \left[ (1 -m_-^2) ( C_+ /C_-
\rho_+^{1-m_+} )^ {m_-^2/1-m_-} - (1-m_+)^2 C_- /C_+ \rho_+
^{m_+^2}\right],\]
which can attain both positive and negative values. A result
similar to (\ref{M1}) can be derived for cylinders in
asymptotically anti de Sitter metric. Here again, $M_1 \leq \frac{1}{4}$ if we demand that
there be no singularity on the axis, that the cosmological constant outside be greater than inside, and that the outer radius of the shell exceeds its inner radius \cite{Bi_Zo}.

By examining expression (\ref{M1}), we find that it is an
increasing function of $C_+$ and $\rho_+$. Since the cylindrical
shell can be regarded as a source of the outer LC metric one
expects the induced linear matter density (\ref{M1}) to increase
with the outer "mass parameter" $m_+$. However, this is the case
only for $m_+$ in specific intervals depending on the radius
$\rho_+$ of the cylinder -- see figure \ref{DiffM1,m+}. For a finite
$\rho_+>0$ and sufficiently small $m_+$, we do find $\partial M_1
/ \partial m_+ >0$, as expected on classical grounds. Notice,
however, that for small $\rho_+$, this "intuitive" behaviour does
not occur even for small $m_+$ (the region around the origin in
figure \ref{DiffM1,m+}).

\begin{figure}[h]
\begin{center}
\epsfbox{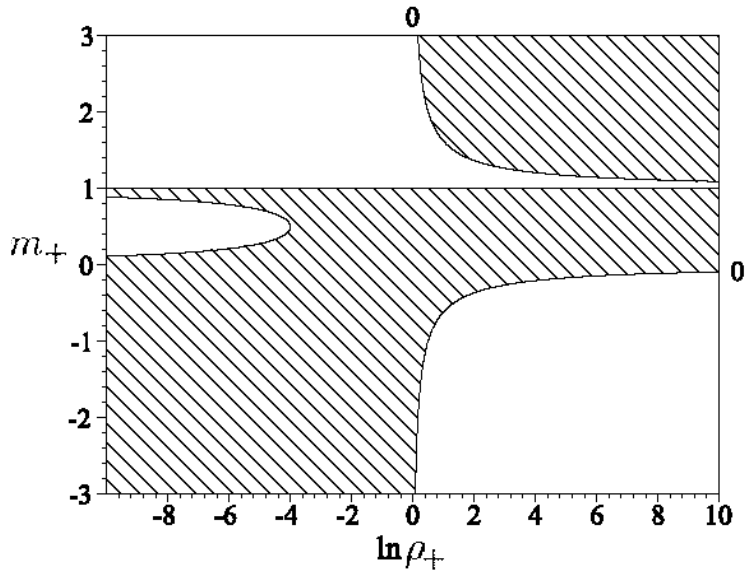}
\end{center}
\caption{\label{DiffM1,m+} The shaded regions indicate the
intervals of $\rho_+$ and $m_+$ in which $\partial M_1 / \partial
m_+>0$, as expected on classical grounds. These regions are given
as follows: $[\rho_+\in (0,e^{-4}] \times m_+ \in
(-\infty,\frac{1}{2}(1-\sqrt{1+\frac{4}{\ln \rho_+}})) \cup
(\frac{1}{2}(1+\sqrt{1+\frac{4}{\ln \rho_+}}),1)] \cup [\rho_+ \in
(e^{-4},1)\times m_+ <1] \cup [\rho_+ \in [1,\infty) \times m_+
\in(\frac{1}{2} (1-\sqrt{1+\frac{4}{\ln \rho_+}}),1)
\cup(\frac{1}{2} (1+\sqrt{1+\frac{4}{\ln \rho_+}}),\infty)]$.}
\end{figure}

Expanding $M_1$ into a series for small $m_+$ we find

\begin{equation}\label{M1-expansion}
\label{expansion} M_1 \sim \frac {1} {4} \left( 1- \frac {1} {C_+}
\right) + \frac {1} {C_+} \frac {m_+} {2} + O(m_+^2).
\end{equation}
The first term represents the contribution from the angular
deficit present in the exterior metric, while the second term
results from the influence of the local curvature of the
spacetime. With no angular defect ($C_+=1$), we obtain the
Newtonian limit as $M_1 \sim m_+/2$. It is, however, clear that
even if $m_+=0$ we can have a non-zero mass per unit length of the
source. This corresponds to a locally flat outer metric with a
missing angle. We only obtain a flat spacetime everywhere if we
put $m_+=0$ and $\rho_- = \rho_+$. Otherwise $C_+ \not = 1$ (
"compensating" the unequal lengths of the hoops in locally flat
inner- and outer spacetimes) and expansion (\ref{expansion}) does
not start with $m_+/2$. This case does not have a Newtonian limit:
one has to consider the global properties of the outer metric
($C_+$) since we are examining total mass per unit length of the
cylinder which is not a local quantity. Our conclusion here is
that Marder's definition \cite{Marder} of mass per unit length
of the source is appropriate in the case of a shell.

Wang et al. constructed some cylindrical shell models \cite{Wang
et al}, \cite{Herrera} for which they discuss (in \S 3 and \S 5) various definitions of the
mass per unit length. In particular, they claim that in the case
of their (somewhat contrived but satisfying energy conditions)
shell of an anisotropic fluid (see their equations (12) and (30), respectively), Marder's
definition gives an incorrect Newtonian limit. To change to their
notation we have to do the following replacements: $C_+
\rightarrow C$, $m_+ \rightarrow 2\sigma$, $\rho_+ \rightarrow
(Ar_0)^{1/A}$ where
\begin{equation} \label{Definition of A}
A= m_+^2-m_++1 = 4\sigma^2-2\sigma+1.
\end{equation}
Now for the shell they consider, we have $C = A^{\frac {1-2 \sigma} {A}}r_0^{\frac
{-4 \sigma^2} {A}} \sim 1-2 \sigma$ which does not approach 1 fast
enough (quadratically in $\sigma$) and thus the parameter $C$
influences the limiting value of $M_1$. Therefore, the actual
value of $M_1$ must be different from $\sigma=m_+/2$ as it follows
from our expansion (\ref{expansion}).
If in any of these cases we do require $C_+=1$, we always obtain
the correct Newtonian limit. For Wang et al., this means $r_0 =
(A(r_0 -a))^{(1-2\sigma)/A}$ (see their equations (5) and (22), respectively) and then $M_1 =
\frac{1}{4} (1- (1-2\sigma)^2 r_0^ {-\frac {4\sigma^2}
{1-2\sigma}})$, which indeed gives $M_1 \sim \sigma +
O(\sigma^2)$. It is essential that for the conicity to be 1 to the second order in $\sigma$, one has to have the radial shift $a$, introduced in references \cite{Wang
et al}, \cite{Herrera}, non-vanishing. This is just the case excluded in \cite{Wang
et al}, \cite{Herrera}.

\begin{figure}[h]
\begin{center}
\epsfbox{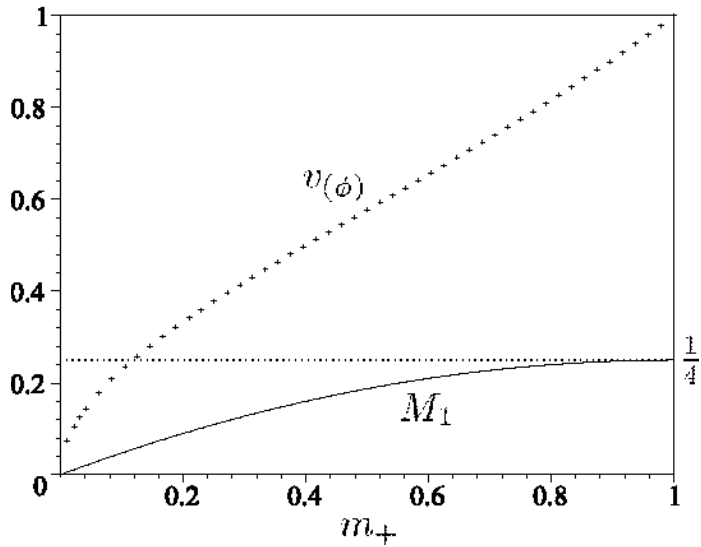}
\end{center}
\caption{\label{Counter-rotating Particles} Static cylindrical
shells of counter-rotating massive particles: the mass per unit
length, $M_1$, and the velocity $v_{(\mathit{\Phi} )}$ of the
particles (measured by static observers) as functions of the
Levi-Civita external-field parameter $m_+$. With increasing $m_+$,
both $M_1$ and $v_{(\mathit{\Phi} )}$ increase monotonically up to
the limiting values $v_{(\mathit{\Phi} )} \rightarrow 1$, $M_1
\rightarrow \frac{1}{4}$.}
\end{figure}

Stachel \cite{Stachel} also considered cylindrical shells, however, he was rather interested in the
physical meaning of the metric parameters than in the structure of
the shell itself. Langer \cite{Langer}, using Israel's formalism,
worked out the case of a cylinder composed of two streams of free particles counter-orbiting in
$\pm\mathit{\Phi}$-directions; and he found agreement of his results
with those of Raychaudhuri and Som \cite{Raychaudhuri} who did not
use the Israel formalism but a limiting procedure starting from
a shell of a finite thickness. It turns out that in this case
the parameter $m_+ \in [0,1)$. The lower limit is due to the condition
of a non-negative matter density whereas the upper one because
geodesics become null for $m_+=1$. Further, we find $C_+ =
\rho_+^{-m_+^2}$ and mass per unit length $M_1 = \frac{1}{4} m_+ (2-m_+)$, independent
of other parameters, is always greater than or equal to $0$. It
has a supremum of $\frac{1}{4}$ at $m_+=1$. The formula for $M_1$ gives the correct weak-field limit.
The velocity of the particles as measured by static observers is
$v_{(\mathit{\Phi})}=\sqrt{\frac {m_+} {2-m_+}}$ (see figure
\ref{Counter-rotating Particles}). If we fix the outer parameters,
the inner radius is also fixed at $\rho_- = C_+^{-\frac {A}
{m_+^2}}$, with $A$ given in equation (\ref{Definition of A}). By letting $C_+$ increase from $0$ to $\infty$,
$\rho_-$ decreases from $\infty$ to $0$.

\begin{figure}[h]
\begin{center}
\epsfbox{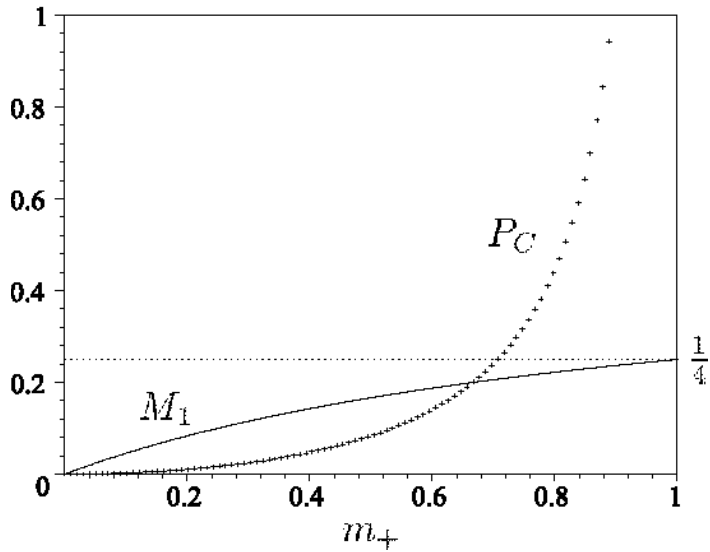}
\end{center}
\caption{\label{Ideal Fluid} Static cylindrical shells of perfect
fluid: the mass per unit length, $M_1$, and the magnitude of the
surface pressure integrated along a ring, $P_C \equiv (2\pi
\rho_-) p$, as functions of the Levi-Civita external-field
parameter $m_+$. With increasing $m_+$, both $M_1$ and $P_C$
increase monotonically up to the limiting values $P_C \rightarrow
+ \infty$, $M_1 \rightarrow \frac{1}{4}$.}
\end{figure}

Langer \cite{Langer} also considered the case of an ideal fluid.
Here $M_1 = \frac{1}{2} \frac {m_+} {m_++1}$ is also independent
of other parameters and it gives the right Newtonian limit.
The admissible range of $m_+$ is $m_+ \in [0,1)$; if we require
$\mu \geq p$ ($\mu$ is the surface energy density), then $m_+ \in
[0,\frac{2}{3}]$. The lower limit is due to the non-negativity of
energy density, the upper one by requiring a finite pressure.
Again, $M_1 \geq 0$, and it has the supremum of $\frac{1}{4}$ for
$m_+=1$. We have $C_+ = (1-m_+^2)/ \rho_+^{m_+^2}$ and $p \equiv
S_{ZZ} = S_{\mathit{\Phi}\mathit{\Phi}} = m_+^2 / 8\pi \rho_-
(1-m_+^2)$. In figure \ref{Ideal Fluid}, the mass per unit length
and the pressure are illustrated. Using the outer parameters, we
can express the inner radius as $\rho_- = C_+^{-\frac {A} {m_+^2}}
(1-m_+^2)^{\frac {1-m_+} {m_+^2}}$, where $A$ is given in equation (\ref{Definition of A}). Increasing $C_+$ means
decreasing $\rho_-$, as in the case of counter-rotating streams.

\section{Generalized models of cylindrical shells}

The shells from counter-rotating particles in
$\pm \mathit{\Phi}$-directions can be generalized in the following
ways:

\begin{enumerate}
\renewcommand{\labelenumi}{(\roman{enumi})}

\item We admit counter-streaming particles to move along $\pm Z$-directions as well ("counter-spiralling motion"). Their velocity can be expressed as $v_{(Z)}^2
\equiv \left( \frac {dZ} {dT} \right)^2= (1-\frac {\rho_+^A}
{\rho_-}) / (\frac {\rho_+^A} {\rho_-} -(1-m_+)^2)$ and
$v_{(\mathit{\Phi})}^2 = m_+^2 / (\frac {\rho_+^A} {\rho_-}
-(1-m_+)^2)$. We find the mass (\ref{M1}) to read

\begin{equation}
M_1 = \frac {1} {4} (1- \frac {\rho_-} {\rho_+^A} (1-m_+)^2).
\end{equation}

Necessary conditions for the velocities $v_{(\mathit{\Phi})},
v_{(Z)}$ to be real read $\rho_+^A/\rho_- \in [A,1]$ and
$(m_+-1)^2 \leq \rho_+^A/\rho_-$. From here we obtain $m_+ \in
[0,1]$ and $M_1 \in [\frac {m_+} {4A},\frac {m_+(2-m_+)} {4}]$ so
that $M_1 \in [0,\frac{1}{4}]$. Again, $M_1 = \frac{1}{4}$
represents the maximum possible mass per unit length. We can
rewrite these conditions in terms of the inner radius as

\begin{equation}
\rho_- \in [C_+^{-\frac {A} {m_+^2}} A^{\frac {1-m_+}
{m_+^2}},C_+^{-\frac {A} {m_+^2}}].
\end{equation}

The lower limit for the inner radius corresponds to the case of
photons\- counter-rotating just in $\pm \mathit{\Phi}$-directions.
The upper limit is given by $\mu, p_Z \geq 0$ and $C_+ =
\rho_+^{1-m_+} / \rho_-$.

\item The shell is composed of photons counter-rotating in $\pm \mathit{\Phi}$-directions (instead of massive particles). We then get $m_+=1$ and,
consequently, $\rho_-=\rho_+=1/C_+$. (In contrast to spherical shells of photons -- see Appendix -- the cylindrical shells can have arbitrary radii depending on $C_+$.) Therefore, $M_1 = \frac {1}
{4}$ and the orbiting velocity is equal to 1.

\item We can also consider counter-streaming photons with each stream being in a spiral motion like massive particles in case (1) above. We find $m_+ \in [0,1]$, $M_1 = \frac {1}{4} \frac {m_+} {A}$. The
angular velocity is $v_{(\mathit{\Phi})}^2 = m_+$ and the velocity
along $\pm Z$-directions reads $v_{(Z)}^2 = 1-m_+$. The inner
radius is at $\rho_- = C_+^{-A / m_+^2} A^{(1-m_+) / m_+^2}$. When
$m_+=1$ this case reduces to the case (ii). In figure
\ref{Spiralling Photons}, the physical velocities and mass $M_1$
are plotted as functions of $m_+$.

\begin{figure}[h]
\begin{center}
\epsfbox{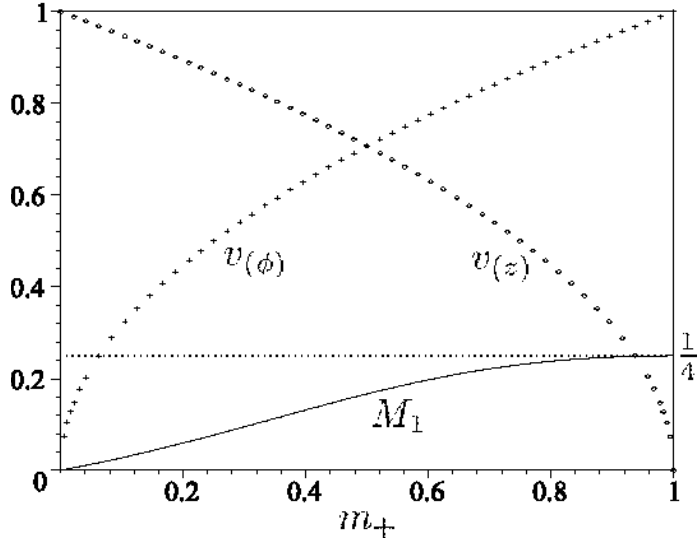}
\end{center}
\caption{\label{Spiralling Photons} Static cylindrical shells of
counter-spiralling photons with physical velocities
$v_{(\mathit{\Phi} )}$, $v_{(Z)}$ ($v_{(\mathit{\Phi}
)}^2+v_{(Z)}^2=1$) as measured by static observers. The mass per
unit length, $M_1$, and both velocities as functions of the
Levi-Civita parameter $m_+$ are illustrated. With $m_+$
increasing, $M_1$ and $v_{(\mathit{\Phi} )}$ monotonically
increase, whereas $v_{(Z)}$ decreases. The limiting point $m_+=1$,
$v_{(\mathit{\Phi} )}=1$, $v_{(Z)}=0$ corresponds to the point
$m_+=1$, $v_{(\mathit{\Phi} )}=1$ in figure \ref{Counter-rotating
Particles}.}
\end{figure}

\end{enumerate}

Let us now investigate a general case of 2-dimensional matter
with a diagonal stress tensor (see equation (\ref{Induced_tensor}))
satisfying various types of energy conditions (see, e.g., \cite{Hawking
and Ellis}, \cite{Wald}).

\begin{enumerate}
\renewcommand{\labelenumi}{\alph{enumi})}
\item   Denoting $S_{TT}=\mu, S_{\mathit{\Phi}\mathit{\Phi}}=p_\mathit{\Phi}, S_{ZZ}=p_Z$, the weak energy
condition, $\mu \geq 0, \mu+p_Z \geq 0, \mu+p_\mathit{\Phi} \geq
0$, for $m_+ \in [0,1]$ implies:

\begin{equation}\label{Weak Energy Condition - lower}
\rho_- \geq C_+^{-\frac {A} {m_+^2}} (1-m_+)^{\frac {2(1-m_+)}
{m_+^2}},
\end{equation}

and for $m_+ \in [1,2]$ we get

\begin{equation}\label{Weak Energy Condition - upper}
\rho_- \leq C_+^{-\frac {A} {m_+^2}} (1-m_+)^{\frac {2(1-m_+)}
{m_+^2}},
\end{equation}

the boundary values of $\rho_-$, corresponding to the equality
signs in (\ref{Weak Energy Condition - lower}) and (\ref{Weak
Energy Condition - upper}), are given by the condition $\mu \geq
0$. We find $\rho_+ \geq (\frac {(1-m_+)^2} {C_+})^\frac {1}
{m_+^2}$, $\frac {\rho_-} {\rho_+^A} \leq \frac {1} {(1-m_+)^2}$
in both intervals of $m_+$.

\item   The strong energy condition requires $\mu+p_Z +p_\mathit{\Phi} \geq 0, \mu+p_Z \geq 0, \mu+p_\mathit{\Phi} \geq
0$. For $m_+ \in [0,\frac{1}{2}]$ this implies that

\begin{equation}\label{Strong Energy Condition - lower}
\rho_- \geq C_+^{-\frac {A} {m_+^2}} (1-2m_+)^{\frac {1-m_+}
{m_+^2}},
\end{equation}

the boundary value of $\rho_-$ is now given by $\mu+p_\mathit{\Phi}
\geq 0$. Here we find $\rho_+ \geq (\frac {1-2m_+} {C_+})^\frac
{1} {m_+^2}$, $\frac {\rho_-} {\rho_+^A} \leq \frac {1} {1-2m_+}$.
For $m_+ \in [\frac{1}{2},2]$, $\rho_-$ and $\rho_+$ can be
arbitrary.

\item   The dominant energy condition, $\mu \geq |p_Z|, \mu \geq |p_\mathit{\Phi}|$,
for $m_+ \in [0,\frac{2}{3}]$ implies that

\begin{equation}\label{Dominant Energy Condition - lower}
\rho_- \geq C_+^{-\frac {A} {m_+^2}} \left[ \frac {m_+^2 -2m_+ +2}
{2} \right]^{\frac {1-m_+}{m_+^2}},
\end{equation}

the boundary value of $\rho_-$ being determined by $\mu \geq p_Z$. We
find $\rho_+ \geq (\frac {m_+^2-2m_++2} {2C_+})^\frac {1}
{m_+^2}$, $\frac {\rho_-} {\rho_+^A} \leq \frac {1}
{1-m_++\frac{m_+^2}{2}}$. For $m_+ \in [\frac{2}{3},1]$ we get

\begin{equation}\label{Dominant Energy Condition - middle}
\rho_- \geq C_+^{-\frac {A} {m_+^2}} (2m_+^2 -2m_+ +1)^{\frac
{1-m_+} {m_+^2}},
\end{equation}

and for $m_+ \in [1,2]$ we finally find

\begin{equation}\label{Dominant Energy Condition - upper}
\rho_- \leq C_+^{-\frac {A} {m_+^2}} (2m_+^2 -2m_+ +1)^{\frac
{1-m_+} {m_+^2}},
\end{equation}

where the boundary value of $\rho_-$ is given by $\mu \geq
p_\mathit{\Phi}$. In the last two intervals we get $\rho_+ \geq
(\frac {2m_+^2-2m_++1} {2C_+})^\frac {1} {m_+^2}$, $\frac {\rho_-}
{\rho_+^A} \leq \frac {1} {1-2m_++2m_+^2}$.

\end{enumerate}

\begin{figure}[h]
\begin{center}
\epsfbox{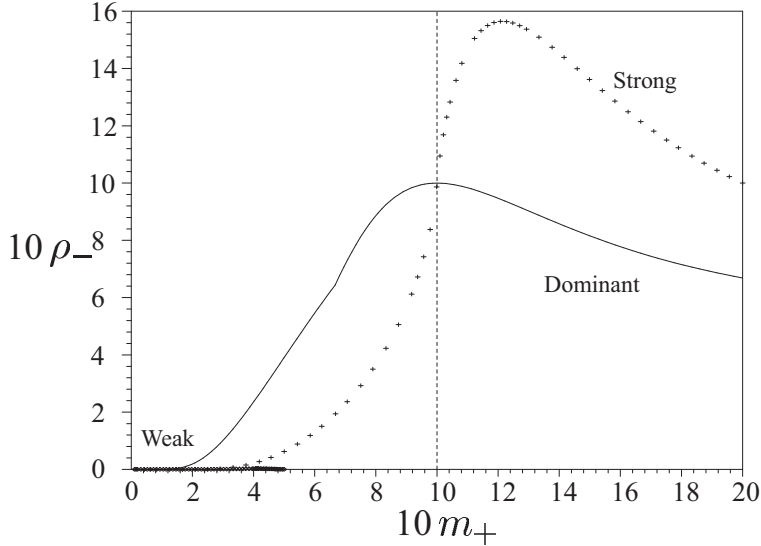}
\end{center}
\caption{\label{Energy Conditions} The radii $\rho_-$ of the
cylinders satisfying the weak, strong and dominant energy
conditions. For $m_+ \in [0,1]$ the admissible range of radii is
{\it above} the corresponding curve. For $m_+ =1$, one has $\rho_-
=1/C_+$. (Here, we fixed $C_+ =1$.) For $m_+ \in
(1,2]$, $\rho_-$ must lay {\it below} the curves. For the details
around the origin, see figure \ref{Energy Conditions - Detail}.}
\end{figure}

\begin{figure}[h]
\begin{center}
\epsfbox{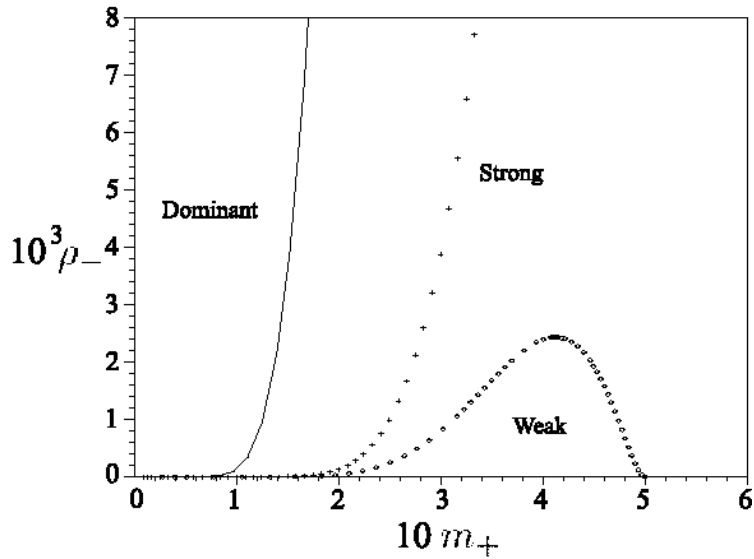}
\end{center}
\caption{\label{Energy Conditions - Detail} Possible radii
$\rho_-$ for small $m_+$ under various energy conditions (see also
the text to figure \ref{Energy Conditions}).}
\end{figure}

Therefore, if $m_+$ is not large ($m_+ \leq \frac {1}{2}$), there
exists a lower limit for possible values of the inner radius
$\rho_-$ of the cylinder in case of all three energy conditions.
In all three cases we allowed $m_+ \in [0,2]$ which follows from
the condition $\mu +p_Z \geq 0$ (in agreement with Wang et al.
\cite{Wang et al}). In figures \ref{Energy Conditions} and
\ref{Energy Conditions - Detail} the above results are illustrated
graphically. Let us also note that if we let $C_+$ change from $\infty$ to $0$,
we find $\rho_-$ changing from $0$ to $\infty$ in all three cases.

What are the limiting values of the mass per unit length of the
cylinders? We can rewrite the expression for the mass as $M_1 =
\frac {1} {4} (1- \frac {\rho_-} {\rho_+^A} (m_+-1)^2)$. From the
above inequalities (\ref{Weak Energy Condition - lower}) --
(\ref{Dominant Energy Condition - upper}) we obtain  $M_1 \in
[0,\frac {1} {4})$ for the weak and dominant energy conditions,
where the lower limit is reached at the maximum possible value of
$\rho_- / \rho_+^A$, whereas the upper bound results from $\rho_-
/\rho_+^A \rightarrow 0$.  In case of the strong energy condition
and $m_+ \in [0,\frac{1}{2})$ we find that $\frac{1}{4} \geq M_1
\geq -\frac {m_+^2} {4(1-2m_+)}$. However, for $m_+ \in
[\frac{1}{2},2]$ there is {\it no} lower bound on $M_1$, the upper
bound remaining the same. The mass diverges to $-\infty$ as
$\rho_+ \rightarrow 0^+$. This can happen since the strong energy
condition does not restrict the value of the density to be
non-negative. It is important that in the interval in which it is
appropriate to speak about a Newtonian limit ($m_+ \in [0,\frac {1}
{2})$) the range of $M_1$ is still restricted from below.

Summarizing, we see that in the above examples the mass per unit
length (equations (\ref{M1}) and (\ref{M1-expansion})) of the cylindrical shell is typically equal neither to
$\frac{1}{4} (1-1/C_+)$ nor to $m_+/2$ but rather it is a
combination of both contributions. As pointed out recently \cite{Anderson}, if we keep the parameters of the outer metric
fixed while decreasing the outer radius $\rho_+$ of the shell, we
inevitably end up with mass per unit length, $M_1$, of the
cylinder diverging, $M_1 \rightarrow -\infty$. The limit $\rho_+
\rightarrow 0$ is permitted only under the strong energy condition
for values of $m_+ \in [\frac{1}{2},2]$, and there is no lower
bound on $M_1$ in this case. Then, however, all other
energy conditions are violated except the strong one.

The only two possible candidates for finite-radius models of a
cosmic string with finite mass per unit length then are: (i) a
shell with $m_+=1, M_1=\frac{1}{4}$, and $\rho_-= \frac {1}
{C_+}$ which satisfies the weak and strong energy conditions
(this is not the case of photons), (ii) a shell with $m_+=0, C_+>1,
M_1=\frac{1}{4}(1-\frac{1}{C_+})$ that satisfies all three
energetic conditions, its circumference is equal to zero.
Indeed, this is a standard model of a cosmic string
\cite{Anderson}, \cite{Linet} -- all its effects are caused by a
deficit angle in a locally flat spacetime.

In Appendix we present corresponding results for classical
cylindrical shells and for relativistic spherical shells.
\\

\section{Test particles outside cylinders}

Let us look at a radial force exerted upon a free particle
at rest in the coordinate system of equation (\ref{Coordinate System}).
The geodesic equation gives

\begin{equation}
\label{radial force}
\frac {d^2 \rho} {d \tau^2} = - \frac {m} {\rho ^{2m^2-2m+1}}.
\end{equation}
The axis is attractive for $m>0$, for $m<0$ it is repulsive. The
"magnitude" of the radial acceleration (the absolute value of
expression (\ref{radial force})) decreases with increasing $\rho$
(and thus also with the proper distance from the axis as this is
an increasing function of $\rho$) for any non-zero $m$. However,
the behaviour due to the changes in $m$ with $\rho$ kept constant
is more subtle (see figure \ref{Diffa,m}). One would expect that a bigger $m$ means a
stronger influence of the centre. This, in general, is not the
case and, moreover, the behaviour depends on the distance from the
axis. For very small distances, $\rho \in (0, \mbox{e}^{-4}]$, and
for $m \in (\frac {1} {4} (1- \sqrt {1+4/\ln \rho}),\frac {1} {4}
(1+ \sqrt {1+4/\ln \rho}))$ the magnitude of the acceleration is
decreasing with increasing $m$. This is against classical intuition
since $m$ is positive in the given interval. For $\rho \in
(\mbox{e}^{-4},1]$ the magnitude is an increasing function of
$|m|$ for any $m$. In the last interval $\rho \in (1, \infty)$ the
acceleration behaves intuitively for $m \in (\frac {1} {4} (1-
\sqrt {1+4/\ln \rho}),\frac {1} {4} (1+ \sqrt {1+4/\ln \rho}))$,
while it is counter-intuitive in the remaining range. In
\cite{Philbin}, the author claims (see p. 1218) that the
acceleration decreases with $m$ for $m \in (\frac{1}{2},1)$
(increasing magnitude). However, this is not the case for
sufficiently large $\rho$. (Also, he relates
the acceleration of particles falling radially from rest to the
existence of circular geodesics, which does not appear to be much telling if we recall, for example, the case of the Schwarzschild metric.)

For any value of $\rho$ there exists an interval of $m$,
containing 0, where the acceleration behaves in accordance with classical intuition. The same
is true for $\rho > \mbox{e}^{-4}$ and $m \in [0,\frac{1}{2}]$.
The first point saves the Newtonian limit, the second provides a
classical intuition for this interval of $m$.

\begin{figure}[h]
\begin{center}
\epsfbox{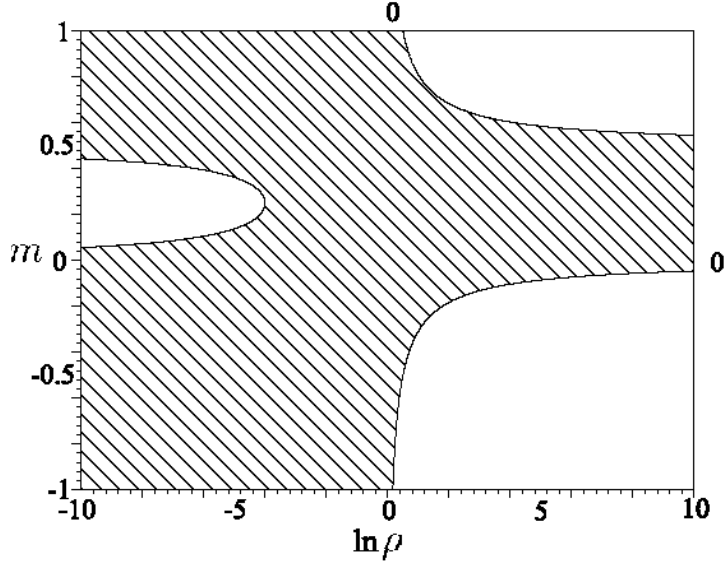}
\end{center}
\caption{\label{Diffa,m} The behaviour of the acceleration of free
test particles at fixed radius $\rho$ under the changes of the
mass parameter $m$. In the shaded regions there is $\partial
|\frac {d^2 \rho} {d \tau^2}| / \partial |m|> 0$.}
\end{figure}

Finally, we study geodesics with $\rho =$ constant. We write
the 4-velocity as $U^ \alpha = U^t (1, v_z, \omega, 0)$. For photons
performing different types of motion we obtain, subsequently

$$\begin{array}{lcl}
z\mbox {-direction} & : & v_z=\pm 1, \;\; \omega=0, \;\; m=2 \;\; \mbox{or} \;\; m=0, \;\; U^t, \rho \;\; \mbox {arbitrary}\\
\varphi\mbox {-direction} & : & v_z=0, \;\; \omega=\pm C, \;\; m=\frac{1}{2}, \;\; U^t, \rho \;\; \mbox {arbitrary}\\
\mbox {spiral motion} & : & v_z= \pm \rho^{m(2-m)} \sqrt {\frac {2m-1} {m^2-1}}, \;\; \omega=\pm C \rho^{2m-1} \sqrt {\frac {m (m-2)} {m^2-1}},\\
&&m \in <0, \frac{1}{2}> \cup <2, \infty), \;\; U^t, \rho \;\;
\mbox {arbitrary}.
\end{array}$$
For particles with non-zero rest mass  we find

$$\begin{array}{lcl}
z\mbox {-direction} & : & v_z= \pm \frac {\rho^{m(2-m)}} {\sqrt {m-1}}, \;\; \omega=0, \;\; U^t= \rho^{-m} \sqrt {\frac {m-1} {m-2}}, \;\; m>2, \;\; \rho \;\; \mbox {arbitrary}\\
&& \mbox {and} \;\; v_z= \pm \sqrt {1-\frac {1} {(U^t)^2}}, \;\; m=0, \;\; U^t \geq 1, \;\; \rho \;\; \mbox {arbitrary}\\
\varphi\mbox {-direction} & : & v_z=0, \;\; \omega=\pm C \rho^{2m-1} \sqrt {\frac {m} {1-m}}, \;\; U^t= \rho^{-m} \sqrt {\frac {m-1} {2m-1}},\\
&&m \in <0, \frac{1}{2}), \;\; \rho \;\; \mbox {arbitrary}\\
\mbox {spiral motion} & : & v_z= \pm [\rho^{m(1-m)} / U^t] [(1-m + (U^t)^2 \rho^{2m}  (2m-1))/(m^2-1)]^{1/2},\\
&& \omega=\pm C [\rho^{m-1} / U^t] [m(1-m +
(U^t)^2 \rho^{2m}  (m-2))/(m^2-1)]^{1/2}.
\end{array}$$
In the last case of spiral motion, the parameters $(m,U^t,\rho)$ are related by the condition that
$v_z$, $\omega$ are real and $v_z^2+\omega^2 <1$. There is a
solution to these inequalities only as long as $m \in (-\infty,
\frac{1}{2}> \cup \; (2, \infty)$. In other words we can choose
two of these parameters and find the admissible range of the
remaining parameter. The admissible intervals of $m$ found above in cases of more general than just circular test particle motion correspond to a possible interchange of the roles of coordinates $\varphi$ and $z$, suggested in \cite{Herrera}.

\appendix
\setcounter{section}{1}
\section*{Appendix}

In this Appendix we summarize some basic properties of Newtonian cylindrical shells and of relativistic spherical shells.\\

A. {\it Newtonian cylindrical shells}

\begin{figure}[h]
\begin{center}
\epsfbox{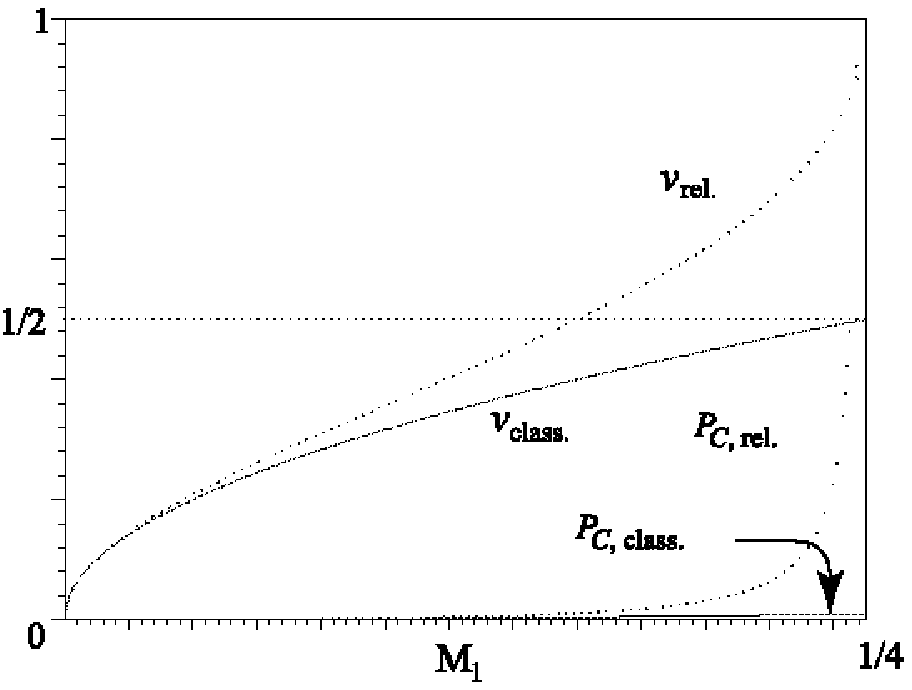}
\end{center}
\caption{\label{Classical Versus Relative} Classical versus
relativistic cylindrical shells. We plot $v_{\mbox{class.}} = r
\omega$, $v_{\mbox{rel.}} = \rho_- \omega$, and 'circumferential'
pressures $P_{C,\:
\mbox{class.}} = (2\pi r)p$ and $P_{C,\:\mbox{rel.}} = (2\pi \rho_-) p$ for the physically appropriate range
of $M_1$.}
\end{figure}

Angular velocity of test particles orbiting a cylinder at
$\rho=\;$constant is $\omega = \frac {\sqrt {2M_1}} {r}$. The
same is true for particles spiralling parallel to the axis. Angular velocity of the particles making up the cylinder of radius
$R$ and mass per unit length $M_1$ is $\omega = \frac {\sqrt {M_1}} {R}$. These particles may
rotate in one direction only and move parallel to the axis as
well. With a relativistic cylinder we have $\omega = \frac
{1} {\rho_-} \sqrt {\frac {1-\sqrt{1-4M_1}} {1+\sqrt{1-4M_1}}}$.
If the thin cylinder consists of an ideal fluid, the pressure
reads $p = \frac {M_1^2} {2\pi r}$. In relativity we get
$p = \frac {M_1^2} {2\pi \rho_- (1-4M_1)}$.
In figure (\ref{Classical Versus Relative}) the behaviour of these characteristic functions is illustrated.\\

B. {\it Relativistic thin spherical shells}

Let the outer Schwarzschild mass parameter be $m$. There
is a flat spacetime inside. The shell is located at $r_-=r_+
\equiv r$. The mass of the
shell is defined as $M \equiv 4\pi r^2 \mu$ with $\mu \equiv
S_{TT}$. We get $S_{TT} = \frac {1-\sqrt{F}} {4\pi r}$, $S_{\mathit{\Theta} \mathit{\Theta}} = \frac {\sqrt{F}-1+\frac {m} {r\sqrt{F}}} {8\pi r}$, and $S_{\mathit{\Phi} \mathit{\Phi}} = \frac {\sqrt{F}-1+\frac {m}
{r\sqrt{F}}} {8\pi r}$, with $ds^2 = -F dt^2 + \frac {dr^2} {F} + r^2 d\theta^2 + r^2 \sin
^2 \! \theta \; d\varphi^2$. The principal pressures are the same due to spherical symmetry. We
require $r \geq 2m$. For $m
\geq 0$ we have $\mu \geq 0$. Regardless of the parameters we
find $p  \equiv S_{\mathit{\Theta} \mathit{\Theta}} =
S_{\mathit{\Phi} \mathit{\Phi}} \geq 0$.

Different energy conditions lead to the following results.

\begin{enumerate}
\renewcommand{\labelenumi}{(\roman{enumi})}

\item Weak energy condition. One gets $m \geq 0$ if $\mu \geq 0$. This is satisfied for any $r \geq 2m$. We find $M_{\mbox{max}} = 2m$ for
$r= 2m$; however, the pressure then diverges.

\item Strong energy condition. One gets $m \geq 0$ if $\mu +2p \geq 0$. This is satisfied for any $r \geq 2m$. The case of $M_{\mbox{max}}$ is the same as in (i).

\item Dominant energy condition. We need $r \geq \frac {25} {12} m$ if $\mu \geq p$. (In this case $m$ may be negative but then also $\mu < 0$ and $p < 0$, with $|p| \geq |\mu|$.) We find $M_{\mbox{max}} = \frac {5} {3}
m$.

\item If the shell is made of photons, we find $r = \frac {9} {4} m$, $M = \frac {3} {2} m$.

\item For a shell of particles with non-zero rest mass we obtain $r > \frac {9} {4} m$ while $M_{\mbox{max}} = \frac {3}{2} m$. For this limiting value of $M$ the trajectories become null.

\end{enumerate}
Maximal $M$ of the shell is again achieved in the case of photons. The dominant energy condition is the most restrictive as in the cylindrical case.

\ack{We thank Ji\v{r}\'\i \; Langer and Tom\'a\v{s} Ledvinka for
discussions. We were supported in part by Grants Nos. GACR 202/02/0735
and GAUK 141/2000 of the Czech Republic and the Charles University.}

\end{document}